\documentclass[acmtog]{acmart}
\acmSubmissionID{223}

\setcitestyle{nosort} 

\usepackage{tikz}
\usepackage{subcaption}
\usepackage{enumitem}
\usepackage{mathtools}

\usepackage{bbm}
\usepackage{booktabs} 
\usepackage[ruled]{algorithm2e} 

\SetAlFnt{\small}
\SetAlCapFnt{\small}
\SetAlCapNameFnt{\small}
\SetAlCapHSkip{0pt}
\allowdisplaybreaks

\def \x {\mathbf{x}}
\def \y {\mathbf{y}}

\def \f {\Delta f}

\def \R {\mathbb{R}^3}

\def \S {\mathcal{S}^2}
\def \B {\mathcal{B}}
\def \A {\mathcal{A}}
\def \w {\omega}
\def \wy {\gamma}
\def \dw {d\sigma}
\def \dl {d\tau}
\def \grz {\x^*}
\def \grzy {\y^*}
\def \n {\mathbf{n}}
\def \v {\mathbf{v}}
\def \vn {\v^{\perp}}
\def \vnw {v^{\perp}}
\def \c {\x}

\def \d {\partial}

\def \epsilon {\varepsilon}
\DeclareMathOperator\sdf{SDF}
\DeclareMathOperator\dsdf{SDF^\prime\!}

\DeclareMathOperator\itx{\textbf{r}}

\def \up {\uparrow}
\def \down {\downarrow}

\setcopyright{none}

\setcopyright{none}
\renewcommand\footnotetextcopyrightpermission[1]{}
\begin{document}

\title{A Simple Approach to Differentiable Rendering of SDFs}

\author{Zichen Wang}
\affiliation{%
 \institution{Cornell University}
 \city{Ithaca}
 \country{USA}}
\email{zw336@cornell.edu}

\author{Xi Deng}
\affiliation{%
 \institution{Cornell University}
 \city{Ithaca}
 \country{USA}}
\email{xd93@cornell.edu}

\author{Ziyi Zhang}
\affiliation{%
 \institution{EPFL}
 \city{Laussane}
 \country{Switzerland}}
\email{ziyi.zhang@epfl.ch}

\author{Wenzel Jakob}
\affiliation{%
 \institution{EPFL}
 \city{Laussane}
 \country{Switzerland}}
\email{wenzel.jakob@epfl.ch}

\author{Steve Marschner}
\affiliation{%
 \institution{Cornell University}
 \city{Ithaca}
 \country{USA}}
\email{srm@cs.cornell.edu}

\begin{teaserfigure}
    \centering
    \includegraphics[width=\linewidth]{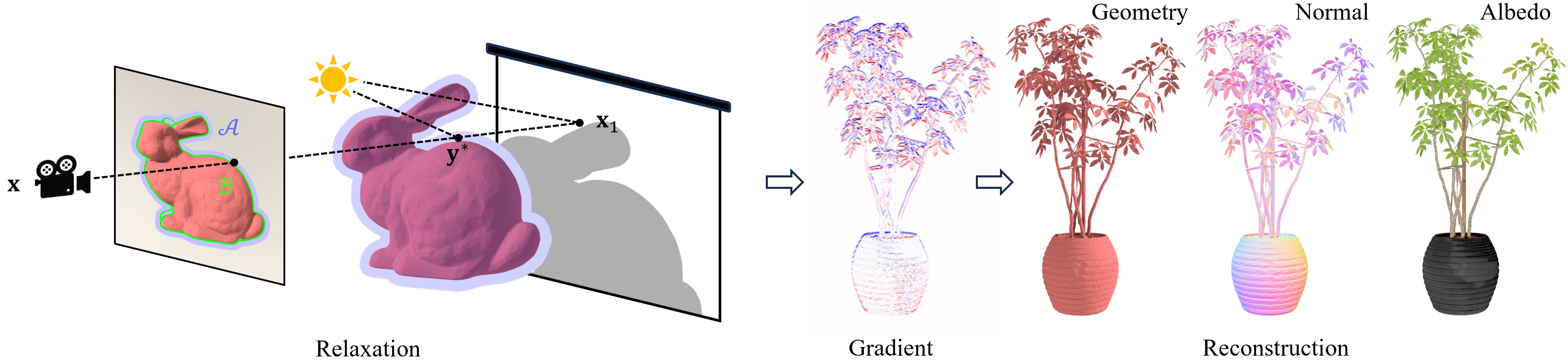}
    \caption{The jump discontinuities at the visibility boundary $\B$ make physically based rendering non-differentiable. To explicitly sample and address such discontinuities, we propose to relax the visibility boundary to a thin band of relaxed boundary $\A$. This corresponds to relaxing the surfaces to thin volumes for sampling discontinuities but retaining surface representations for forward rendering. Consequently, we are able to compute accurate gradients of the rendered images with respect to the scene parameters and, in turn, achieve high-quality inverse rendering results, with less complexity than previous methods.}
    \label{fig:teaser}
\end{teaserfigure}

\begin{abstract}

We present a simple algorithm for differentiable rendering of surfaces represented by Signed Distance Fields (SDF), which makes it easy to integrate rendering into gradient-based optimization pipelines. To tackle visibility-related derivatives that make rendering non-differentiable, existing physically based differentiable rendering methods often rely on elaborate guiding data structures or reparameterization with a global impact on variance. In this article, we investigate an alternative that embraces nonzero bias in exchange for low variance and architectural simplicity. Our method expands the lower-dimensional boundary integral into a thin band that is easy to sample when the underlying surface is represented by an SDF. We demonstrate the performance and robustness of our formulation in end-to-end inverse rendering tasks, where it obtains results that are competitive with or superior to existing work.
    
\end{abstract}

\maketitle


\section{introduction}

Gradient-based methods have shown remarkable success in optimization problems that are often associated with high-dimensional parameter spaces. Effectively backpropagating gradients requires each step of a computation to be differentiable. Unfortunately, this is by default not the case for physically based rendering methods, where visibility discontinuities arise from boundaries of visible regions (e.g. silhouette and self-occlusions). Naive automatic differentiation normally builds on the assumption that the derivative of an integral matches the integral of a derivative. However, the influence of geometric parameters on discontinuous regions of the integrand sadly breaks this important relationship, which causes the computed derivatives to be so severely biased that they generally cannot be used.

A number of prior works have proposed solutions to this problem. They broadly fall into two categories: \emph{boundary sampling} methods \cite{Li:2018:DMC, Zhang:2020:PSDR} evaluate a lower-dimensional boundary integral to remove bias, often with complex data structures to help sample the boundaries; \emph{area sampling} methods \cite{Loubet2019Reparameterizing, bangaru2020warpedsampling, Vicini2022sdf, Bangaru2022NeuralSDFReparam} leverage reparameterization or the divergence theorem to convert the boundary into a finite region, usually at the cost of significantly increased gradient variance. Following this classification, our newly proposed method blends the two classes of methods to achieve both architectural simplicity and low gradient variance. 

The idea of our method is simple: we define a narrow finite band near the boundary and extend the boundary integrand over that region.  We call this \emph{relaxation} because it relaxes the condition defining the visibility boundary (that paths exactly graze a surface in the scene) to a looser condition (that they come near a surface in the scene).  We show that by defining this relaxation in the right way, we can easily compute the required integrand with minimal additional machinery.  We demonstrate that this method is competitive in terms of total error with more complex existing unbiased methods, and that it is efficient and robust enough to be applied in practical gradient-based pipelines, such as for reconstructing complex geometry. 


\section{related works}

\paragraph{Differentiable rasterization}

Several works \cite{liu2019softras, liu2020dist, loper2014opendr, cole2021differentiable} propose to blur the silhouettes of triangle meshes into a probabilistic distribution, or to smooth the rasterized image to make the rendering process directly differentiable. Our boundary relaxation bears similarities to the use of blur in these methods. More recently, NVDiffRast \cite{Laine2020diffrast} realizes a family of lower-level primitive operations that compose into a complete differentiable rasterization pipeline that performs analytic post-process antialiasing to handle boundaries. In general, differentiable rasterization can be made highly efficient but cannot effectively model higher-order transport and scattering effects.

\paragraph{Physically based differentiable rendering}

Methods in this category rely on the Monte Carlo method to faithfully reproduce the desired physical phenomena. The main challenge is that boundary discontinuities and self-occlusions interfere with the differentiation of the underlying integrals. To address the resulting bias, Li et. al.~\shortcite{Li:2018:DMC} compute a separate \emph{boundary integral} at each shading point, which they sample using a 6D Hough tree. Path Space Differentiable Rendering (PSDR)~\cite{Zhang:2020:PSDR} builds light paths ``from the middle,'' by sampling a path segment tangent to a mesh edge and performing bidirectional random walks to turn them into a full path.
 
Area sampling methods are based on the idea of \emph{reparameterizing} integrals with different coordinates, whose derivative with respect to scene parameters smoothly interpolates the motion of boundaries~\cite{Loubet2019Reparameterizing, bangaru2020warpedsampling}. In particular, researchers have  proposed specialized parameter constructions for SDFs~\cite{Vicini2022sdf, Bangaru2022NeuralSDFReparam}, which enable easy identification of rays that pass close to the scene geometry. Our proposed method uses SDFs for the same reason. 

Recently, methods have been proposed as a blend of the two sampling methods. Projective sampling \cite{Zhang2023Projective} collects paths that are close to the boundary during forward rendering and projects them to the boundary. Similar to projective sampling, we also collect near-silhouette paths, but we use them to directly approximate the boundary integral. This enables us to retain the performance of area sampling methods without requiring the construction of smooth warp fields. At the same time, we do not need additional sampling steps or acceleration data structures common in boundary sampling methods. While our method introduces bias, we show that this bias is small enough not to impact convergence when solving inverse problems.

Shape representation is another perspective from which to classify differentiable rendering methods. Mesh-based methods \cite{Li:2018:DMC, Nicolet2021Large} are fast for ray intersection but are non-smooth and awkward for shape optimization. Implicit surfaces like SDFs \cite{Vicini2022sdf, Bangaru2022NeuralSDFReparam} are a better-behaved parameterization for shape, produce smooth surfaces, and support distance queries easily, but tend to be slower. Some methods use a hybrid \cite{remelli2020meshsdf, Cai:2022:PSDR-MeshSDF, mehta2022level}, employing a differentiable surface extraction method that converts an implicit surface to a mesh. Our method is not fundamentally tied to a particular representation, but we use SDFs because their smoothness is convenient and they easily support the type of distance query we need.

\paragraph{Applications of differentiable rendering}

An important application of a differentiable render is for \textit{3D reconstruction}. In 2020, Yariv et. al. \shortcite{yariv2020multiview} proposed an Implicit Differentiable Renderer (IDR) for object-centric reconstruction. Later, researchers applied differentiable rendering as a post-processing step to refine the reconstructions obtained by other methods \cite{iron-2022, sun2023neuralpbir}. The necessity of \textit{surface representations} constitutes one of the main bottlenecks to the performance of physically based differentiable rendering. Jump discontinuities inevitably arise when rays cross the surface silhouette and intersect with different surfaces, requiring either discontinuity handling or additional mask supervision \cite{yariv2020multiview}.

Existing methods in this field mainly adopt \textit{volume representations}, such as radiance fields \cite{mildenhall2020nerf} or Gaussian Splats \cite{kerbl3Dgaussians}. One of the advantages of such volume representations is that volume rendering is fully differentiable. To extract the underlying surface, we can either apply ad-hoc mesh extraction \cite{rakotosaona2023nerfmeshing, yariv2023bakedsdf,  tang2022nerf2mesh} or jointly train an SDF network \cite{wang2021neus, yariv2021volume, li2023neuralangelo}. All of these work under the premise that the volume converges near the target surface. From this perspective, we can say that volume representations relax the entire surface for differentiability, while our method relaxes the surface partially: we retain a surface representation for forward rendering but relax the surfaces to a thin volume for discontinuity sampling.

Surface representations and physically based differentiable rendering are more widely adopted in \textit{inverse rendering} tasks, which seek to jointly reconstruct the geometry, material, lighting, etc. \cite{physg2021, iron-2022, zhang2022invrender, verbin2023eclipse}. These physical quantities often require the simulation of full light transport. Finally, differentiable rendering is also widely used in generative AI \cite{cole2021differentiable, lin2023magic3d}, sensor design \cite{hazineh2022dflat}, and visual arts \cite{article}. 



\section{Method}

\begin{figure} 
    \centering
    \includegraphics[width=\linewidth]{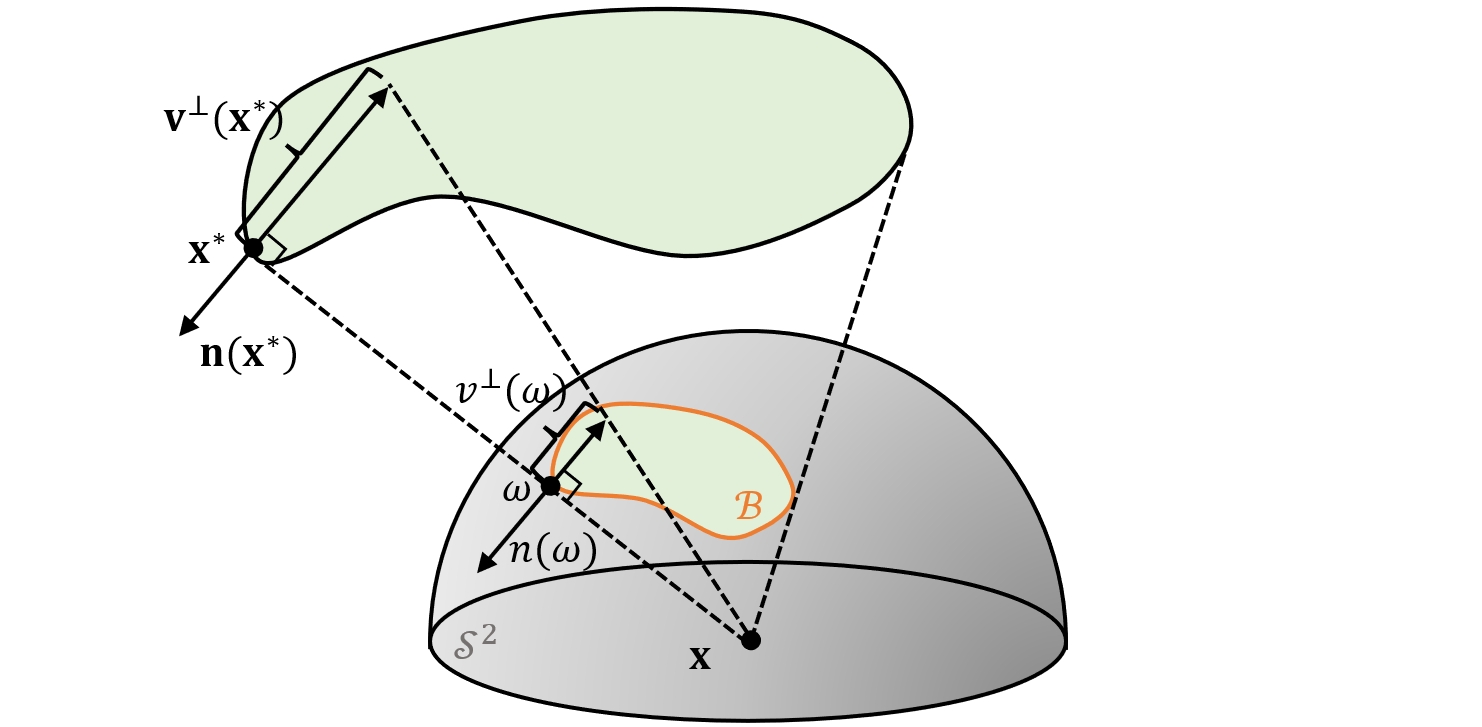}
    \caption{\textbf{Normal velocity}. Each direction $\w$ in the visibility boundary $\B$ corresponds to a distant silhouette point $\grz$, where the associated ray grazes an occluder. The dot products of the velocity and the normal give the normal velocities in each space, which are inversely proportional to $||\x-\grz||$.}
    \label{fig:velocity}
\end{figure}

\subsection{Preliminaries: Differentiating the Rendering Equation}

The \textit{rendering equation} states that the outgoing radiance $L_o$ at a point $\x$ in direction $\w$ is
\begin{equation}
    L_o(\x, \w_o) = L_e(\x,\w_o) + \int_{\S} f_s(\x, \w, \w_o) L_i^\perp(\x, \w) 
                       ~\dw(\w),
\end{equation}
where $\w\in\S$ is a vector on the unit sphere, $L_e$ models emission, $\dw$ denotes the solid angle measure, $f_s$ is the BSDF, and $L_i^\perp(\x,\w) = L_o(\itx(\x, \w),-\w) \lvert\langle \w,\n(\x) \rangle\rvert$, where $\mathbf{r}$ is the {ray intersection} function.  We assume that $L_e$ and $f_s$ are smooth so that discontinuities in the integrand only arise due to visibility changes at object boundaries. In the following, we abbreviate the above to 
\begin{equation}
    \label{eq:render}
    I = \int_{\S} f(\w) ~\dw(\w).
\end{equation}

Our goal is to compute the derivative $\d I/\d\theta$ with respect to a scene parameter $\theta$ that potentially influences the placement of discontinuities. Previous work \cite{Zhang:2020:PSDR,bangaru2020warpedsampling} observed that this derivative can be expressed as a sum of two integrals:
\begin{equation}
    \label{eq:li-two-interior-and-boundary}
    \frac{\d I}{\d\theta} = \int_{\S} \frac{\d f(\w)}{\d\theta} ~\dw(\w) 
                          + \int_{\B} \vnw(\w)\f(\w) ~\dl(\w).
\end{equation}
These two terms are the \textit{interior} and \textit{boundary} integrals. In the latter term, $\dl(\w)$ denotes the arclength measure along $\B\subset\S$, the set of visibility-induced discontinuities observed at $\x$. Each direction $\w\in\B$ on such a boundary is associated with a distant silhouette point that we label $\grz$. The function $\Delta f(\w)$ equals the step change in incident radiance across this oriented boundary.

The Signed Distance Function $\sdf: \R\to\mathbb{R}$ describes the distance of a point $\x$ to the implicit surface $\sdf^{-1}(0)$. By convention this distance is negative inside and positive outside the implicit surface. Such definition over the entire scene space enables us to define a normal field $\n(\x) \coloneqq \nabla\sdf(\x)/\|\nabla\sdf(\x)\|$ that smoothly extends to positions $\x\in\R$ in the neighborhood of the surface. The \emph{normal velocity} $\v(\x) \coloneqq d\x/d\theta$ is defined as the change of the surface at $\x$ along its normal with respect to a perturbation of $\theta$. It equals the following normal-aligned vector field \cite{normalvel}:
\begin{equation}
    \v(\x) = - \frac{\d}{\d\theta}\sdf(\x) \cdot \frac{\nabla\sdf(\x)}{\|\nabla\sdf(\x)\|^2}.
\end{equation}
The \textit{scalar normal velocity} $\vn(\x)$ is a key quantity that measures the projection of this velocity $\v(\x)$ onto the normal $\n$:
\begin{equation}
\vn(\x) := \langle \v(\x), \n(\x) \rangle.
\end{equation}

Each of these terms has its spherical projection. The velocity of a direction $\w\in\S$ is denoted $v(\w) \coloneqq d\w/d\theta$. For $\w\in\B$, we can further define the \textit{boundary normal} $n(\w)$ to be the outward-pointing unit tangent vector perpendicular to $\B$. The scalar normal velocity is then $\vnw(\w) \coloneqq \langle v(\w), n(\w) \rangle$. Note that we use \textbf{bold} notation for variables in $\R$ and \textit{italic} for corresponding variables on $\S$.

To relate the spatial scalar normal velocity with its spherical projection, we observe that $\w\in\B$ and its corresponding silhouette point $\grz$ satisfies
\begin{equation}
    \grz = \x + \|\grz-\x\|~\w.
\end{equation}
Taking its derivative with respect to $\theta$ we have
\begin{equation}
    \v(\grz) = \frac{\d}{\d\theta}\|\grz-\c\|\cdot\w + \|\grz-\c\|\, v(\w).
\end{equation}
Taking an inner product with respect to $n(\omega)$ on both sides yields
\begin{equation}
    \label{eq:vn}
    \vn(\grz) = \|\grz-\c\|~\vnw(\w).
\end{equation}
That is, the normal motion on the unit sphere is inversely proportional to the distance between $\x$ and $\grz$ (\autoref{fig:velocity}).

\begin{figure}
    \centering
    \includegraphics[width=\linewidth]{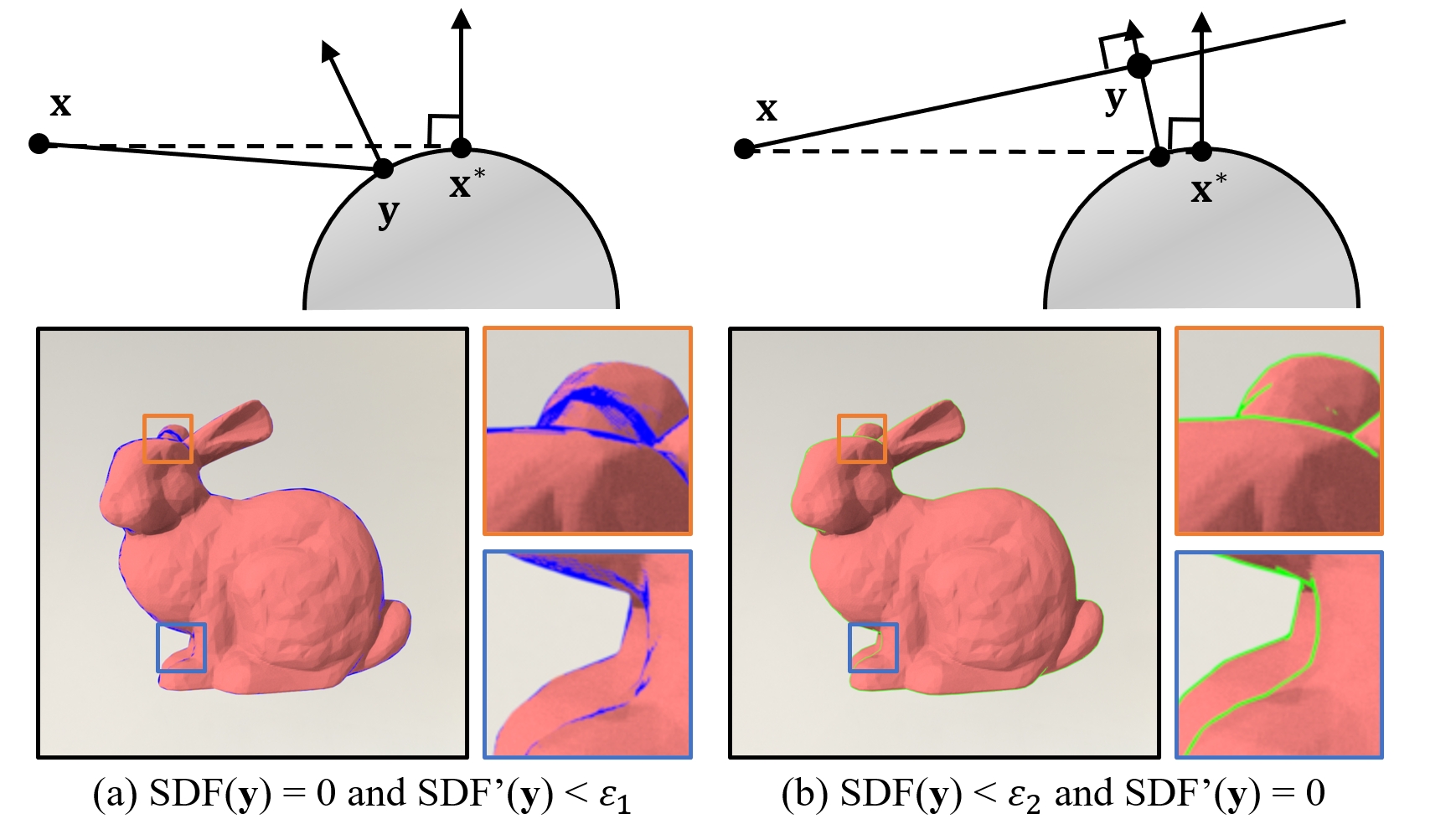}
    \caption{\textbf{Relaxation}. We color image plane samples whose corresponding $\y$s satisfy the relaxation conditions blue/green to visualize the sampling of the silhouette after relaxation. (a) Relaxing the directional derivative condition leads to rays that intersect with the surface almost tangentially. (b) Relaxing the SDF condition leads to rays that graze the surface with no intersection. Here we set $\epsilon_1 = 0.2$ and $\epsilon_2 = 0.002$. We see that relaxing the SDF condition leads to more uniform samples around the silhouette.}
    \label{fig:relax}
\end{figure}

\subsection{Boundary Relaxation}

A silhouette point $\grz$ necessarily satisfies
\begin{equation}\label{eq:grz-exact}
    \sdf(\grz)=0 \text{ and } \dsdf(\grz)=0
\end{equation}
where $\dsdf(\x)$ denotes the directional derivative of $\sdf(\x)$ along the ray direction at $\x$ \cite{bangaru2020warpedsampling, Gargallo2007}. $\sdf(\grz) = 0$ requires $\grz$ to be on the surface, while $\dsdf(\grz) = 0$ says that $\grz$ is a local SDF minimum along the ray ($\grz$ cannot be a local maximum since $\sdf$ is non-negative for any points on a valid path segment.) In other words, a silhouette point corresponds to a ray that tangentially intersects with a surface.

Since directly sampling the lower-dimensional silhouette is difficult, a natural idea is to relax the conditions and use nearby rays to approximate the silhouette. In Figure \ref{fig:relax}, we compare between relaxing different conditions. If we relax the $\dsdf$ condition, we will obtain rays that intersect almost tangentially with a surface. If we relax the $\sdf$ condition, we will obtain rays that graze the surface with no intersection. Although relaxing $\dsdf$ seems to give ray intersection points that resemble silhouette points, they are in practice equally hard to sample and sensitive to the curvature at the silhouette. In fact, several recent methods in this direction need to take a large relaxation (sometimes the entire surface) and then take an extra step to walk ray intersection points to the silhouette \cite{iron-2022, Zhang2023Projective}. On the other hand, if we want to use these relaxed points to directly approximate the boundary integral, it is more desirable to relax the condition on the $\sdf$ to 
\begin{equation} \label{eq:grz}
    0 < \sdf(\grzy) < \epsilon \text{ and } \dsdf(\grzy)=0
\end{equation}
for some small $\epsilon > 0$. We call $\epsilon$ the \textit{SDF threshold}. We call $\grzy$ a \textit{relaxed silhouette point}. We call the set of directions $\gamma$ that correspond to these relaxed silhouette points the \textit{relaxed boundary} $\A$. Intuitively, $\A$ is a thin band on the unit sphere on one side of $\B$.

\begin{figure}
    \centering
    \includegraphics[width=\linewidth]{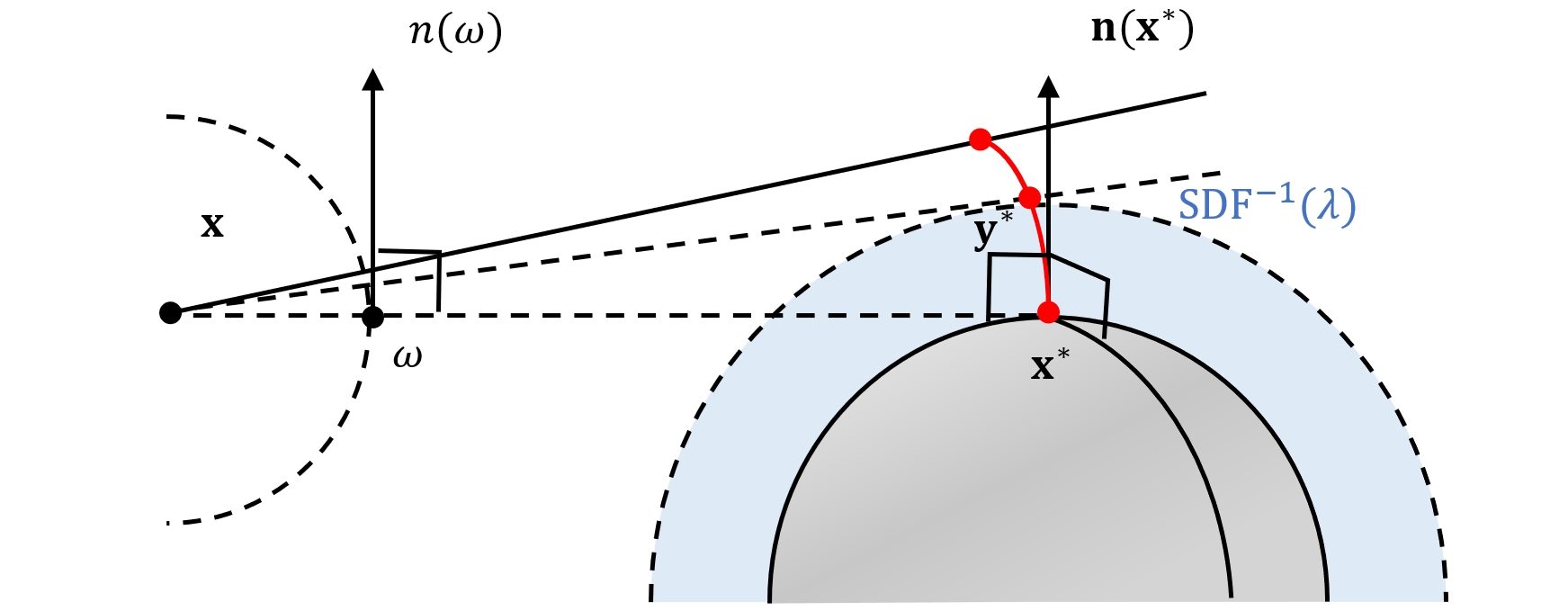}
    \caption{\textbf{Relaxed silhouette point}. Extending along the normal of a silhouette point $\grz$ gives a set of relaxed silhouette points $\grzy$ (red curve) that can be approximated to first order by a line segment. A relaxed silhouette point $\grzy$ can be seen as a silhouette point of the $\lambda$-level set, where $\lambda = \sdf(\grzy)$.}
    \label{fig:relaxed_silhouette}
\end{figure}

Under this definition, there is a natural extension of the boundary integrand from the visibility boundary $\B$ to the relaxed boundary $\A$. Given some direction $\wy\in\A$ and its corresponding relaxed silhouette point $\grzy$, the trick is to see $\grzy$ as a silhouette point of the $\lambda$-level set, where $\lambda = \sdf(\grzy)$ (\autoref{fig:relaxed_silhouette}). In this way, we can apply \autoref{eq:vn} to talk about the scalar normal velocity of $\wy$, and $\f(\wy)$ should be the difference of $f(\wy)$ between missing and intersecting with the $\lambda$-level set. 

Given some $\w\in\B$, we also want to know the set of $\wy\in\A$ that $\w$ relaxes to. One natural thought is to extend $\w$ along its normal direction $n(\w)$. This would correspond to extending $\grz$ along the normal direction $\n(\grz)$ until the relaxed silhouette point $\grzy$ along the ray achieves maximum relaxation $\sdf(\grzy) = \epsilon$. Since $n(\w)$ is identical to $\n(\grz)$ and $\epsilon$ is very small, this extension approximately gives us a segment along $\n(\grz)$ of length $\epsilon$ and, scaled by the distance, a segment along $n(\w)$ of length 
\begin{equation}
    \label{eq:l}
    l(\w) = \epsilon/r
\end{equation}
where $r \coloneqq \|\grz - \c\|$.  We call $l(\w)$ the width of the band. This approximation by a small segment along the tangential direction is then a first-order approximation.


\subsection{Estimating the Boundary Integral}

Now that we know how to relax the lower-dimensional visibility boundary $\B$ to the relaxed boundary $\A$, it remains to ask how can we integrate over $\A$ to estimate the boundary integral
\begin{equation}
    I_{\B} = \int_{\B} \vnw(\w)\f(\w) ~\dl(\w)
\end{equation}

Given how normal extension relates silhouette points and relaxed silhouette points, a good starting point is to weigh the integrand by the width of the band
\begin{equation}
    I_{\B} \approx I_{\A} = \int_{\A} \frac{1}{l(\w)} \cdot \vnw(\w)\f(\w) ~\dw(\w).
    \label{eq:relbdy1}
\end{equation}
As discussed above, the integrand is well-defined as we can see $\grzy$ as the silhouette point of the $\lambda$-level set. Of course, this will only work if the integrand is continuous near the visibility boundary, which is true as long as the path segment is not tangential to multiple surfaces, which we assume is rare, and the surface BRDF is a continuous function of both position and direction. In general, it is not uncommon to regularize the appearance by spatial displacements \cite{yu_and_fridovichkeil2021plenoxels, rosu2023permutosdf}. 

Finally, substituting Equation \ref{eq:vn} and \ref{eq:l} into \autoref{eq:relbdy1} we come to the central result of this paper, which we call the \textbf{\textit{relaxed boundary integral}}:
\begin{align}
    I_{\A} &= \int_{\A} \frac{r}{\epsilon} \cdot \frac{1}{r}\vn(\grzy) \f(\w) ~\dw(\w) \\
           &=       \frac{1}{\epsilon} \int_{\A} \vn(\grzy)\f(\w) ~\dw(\w) \\
           &=       \frac{1}{\epsilon} \int_{\S} \vn(\grzy)\f(\w)\mathbbm{1}_{\A}(\w) ~\dw(\w).
\end{align}

A few remarks on the implications of this new integral: First, the relaxed boundary integral brings the integral domain from a lower-dimensional curve back to the entire unit sphere. This means that we can now re-use the samples for forward rendering to estimate the boundary integral, without additional projection, guiding, or data structures to help sample the silhouette. It follows that many importance sampling strategies used to estimate the forward rendering integrand $f(\w)$ can now also benefit the estimate of $\f(\w)$. Second, the relaxed boundary integral requires minimal additional machinery. Apart from the scalar normal velocity $\vn(\grzy)$ and the difference term $\f(\w)$ inherited from the original boundary integral, we only need to check if a direction is inside the relaxed boundary $\mathbbm{1}_{\A}(\w)$, i.e., whether there is a point satisfying \autoref{eq:grz} when tracing a ray along this direction. This boils down to simply finding the minimal SDF values along the ray (see Section \ref{sec:bisection}).

\begin{algorithm}[t]

\SetAlgoNoLine
\SetKwFunction{Lo}{$L_o$}
\SetKwProg{Fn}{Function}{:}{}

\Fn{\Lo{$\x, \w_o$}}{ 
    // normal path tracing \\
    $\w_i$ = sample an incoming direction \\
    $p$ = evaluate the PDF of $\w_i$ at $\x$ \\
    $f_s$ = evaluate the BSDF from $\w_o$ to $\w_i$ at $\x$ \\
    $\x_1 = \itx(o=\x, d=\w_i)$ // ray intersection \\
    $L_1 = f_s \cdot L_o(\x_1, -\w_i) / p$ // estimated outgoing radiance \\
    // differentiable rendering \\
    \uIf{there exists $\grzy \in \overline{\x\x_1}$}{
        $L_2 = f_s \cdot L_o(\grzy, -\w_i) / p$ \\
        $\n(\grzy), \v(\grzy)$ = evaluate the normal and velocity at $\grzy$ \\
        $dL = \d L/\d\theta + \langle \n(\grzy),\v(\grzy) \rangle (L_2 - L_1) / \epsilon$
    } 
    \Else{
        $dL = \d L/\d\theta$ // interior only
    }
    \KwRet{$L_1$, $dL$}
}

\caption{Differentiable Renderer}
\label{algo}

\end{algorithm}






\begin{figure*}
    \centering
    \includegraphics[width=\linewidth]{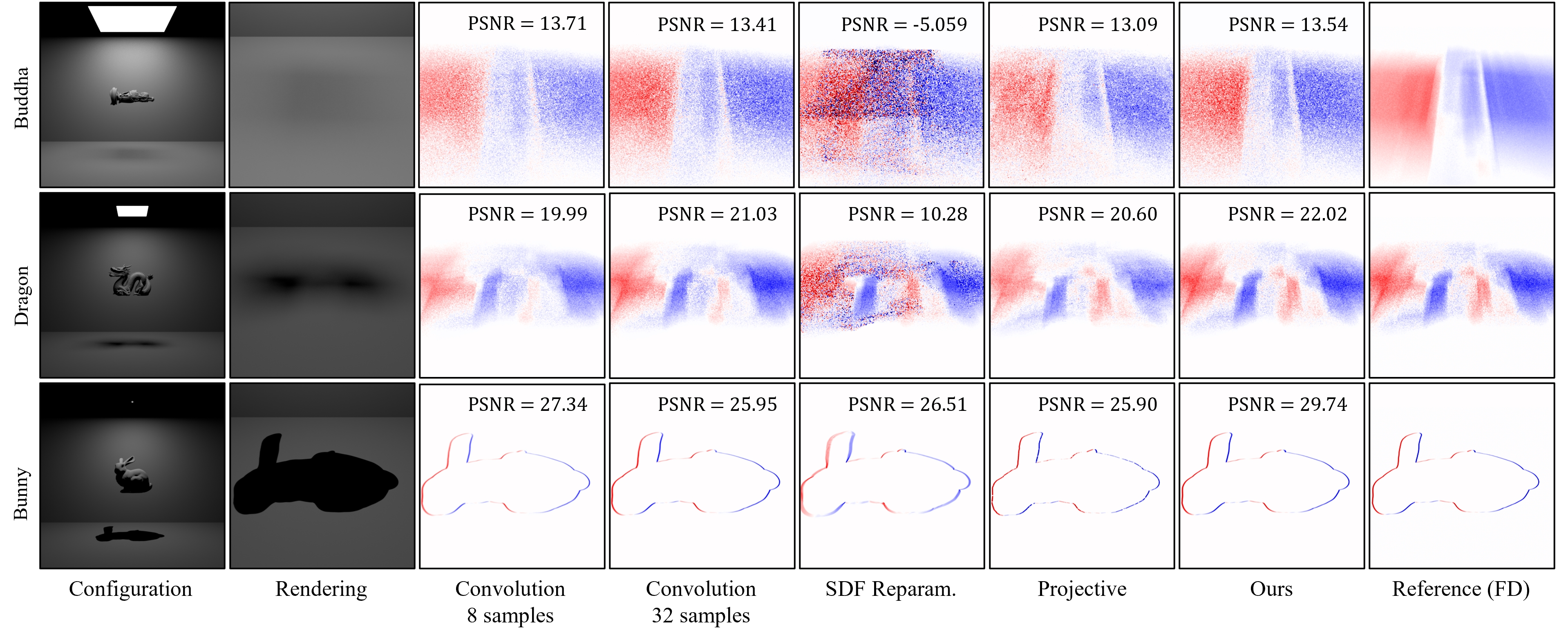}
    \caption{\textbf{Forward derivative}. We use forward-mode differentiation to compute the derivative of the rendered image with respect to a translation along the $x$ axis. We test different shapes under different sizes of square area light and compare with SDF convolution \cite{bangaru2020warpedsampling}, SDF Reparameterization \cite{Vicini2022sdf}, and mesh Projective Sampling \cite{Zhang2023Projective}. For all tests we use a direct integrator with $512\times512$ resolution and $1024$ spp.}
    \label{fig:derivative}
\end{figure*}




\section{Implementation}

\paragraph{Solving for the relaxed silhouette point}
\label{sec:bisection}

While sampling the silhouette is difficult, sampling the relaxed silhouette is much easier. During sphere tracing, we can compute the directional derivative at the intermediate steps and, together with the SDF value, check if we have passed a minimal point. Specifically, if in the previous step the directional derivative is negative and at this step the derivative becomes positive, then this would be a signal that we are near a local minimum. Among all such intermediate steps, we choose the one with the smallest SDF value as our initial guess and run the bisection method for $12$ iterations to pinpoint the relaxed silhouette point.

\paragraph{Estimating relaxed boundary integral}

In \autoref{algo}, we give the pseudo code for our differentiable renderer. In addition to normal path tracing, if there exists a relaxed silhouette point $\grzy$ during the sphere tracing process, we need to re-evaluate the shading at $\grzy$. Taking its difference with the shading at next intersection point gives us the delta difference term. Here we only write down BSDF sampling for simplicity, but our actual implementation leverages multiple importance sampling (MIS). For shadow rays in emitter sampling, note one side of the visibility boundary is always occluded, so we no longer need to evaluate the shading at the relaxed silhouette point. We estimate the interior integral in \autoref{eq:li-two-interior-and-boundary} using automatic differentiation except for the ray intersection process, whose derivatives are computed analytically after the sphere tracing process in the same way as in \cite{Vicini2022sdf, Bangaru2022NeuralSDFReparam}.

\begin{figure}
    \centering
    \includegraphics[width=\linewidth]{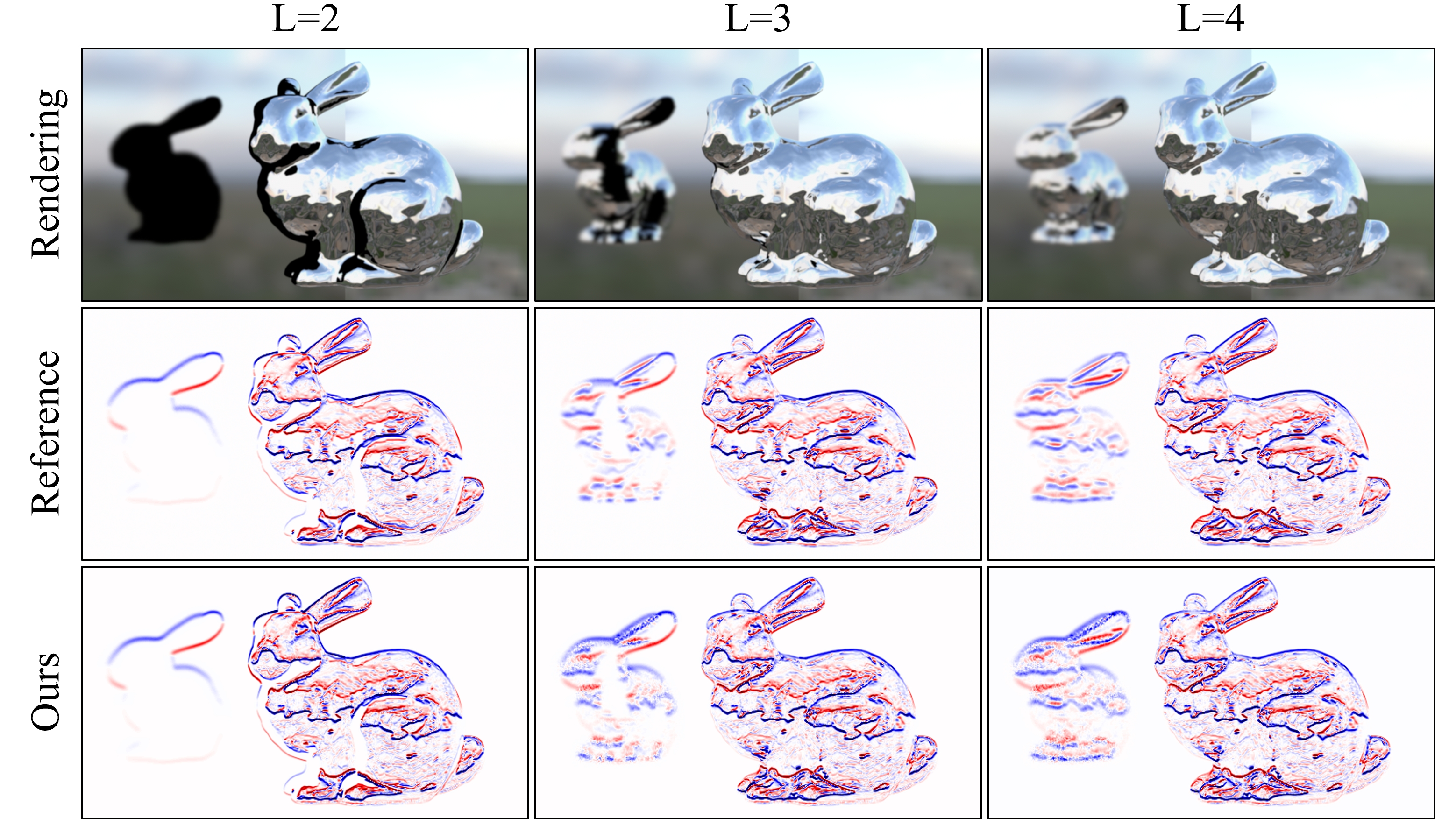}
    \caption{\textbf{Higher-order light transport}. We place a metallic bunny in front of a mirror to validate our differentiable renderer on higher-order light transport. In the first row, we show our rendering under different maximum depths. In the second and third row, we compute the forward derivatives with respect to a translation in the $y$ axis.}
    \label{fig:mirror}
\end{figure}




\begin{figure*}[h!]
    \centering
    \includegraphics[width=\linewidth]{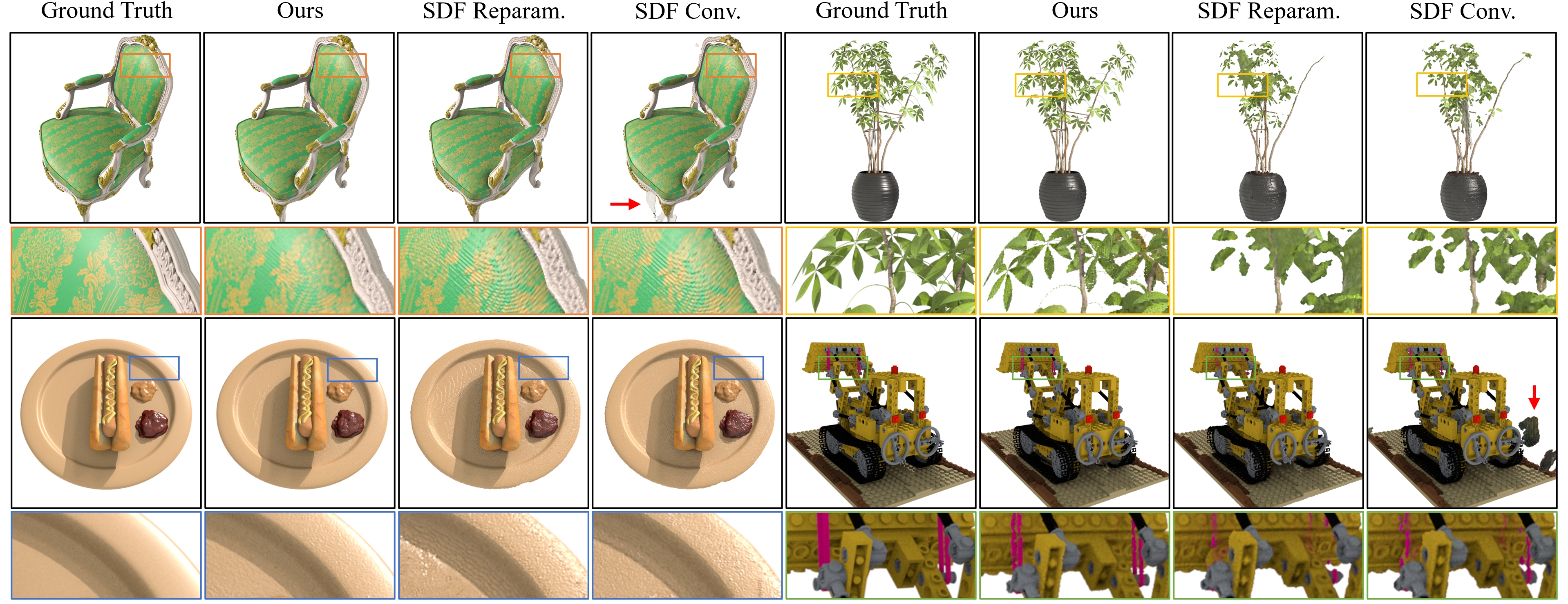}
    \caption{\textbf{Side-by-side comparison}. We compare the final inverse rendering of SDF Convolution (8 auxiliary spp), SDF Reparameterization, and our method using the same optimization setup: in total we run $5000$ iterations optimizing $50$ views; in each iteration, we optimize a batch of $5$ views, rendered with $512\times512$ resolution and $64$ spp. In all test cases, our method results in comparable or more accurate reconstructions.}
    \label{fig:compare}
\end{figure*}

\begin{table*} 
    \centering
    \begin{tabular}{cccccccccccc}
        \toprule
        & \multicolumn{3}{c}{\textbf{Novel Views}} & \multicolumn{3}{c}{\textbf{Relighting (High)}} & \multicolumn{3}{c}{\textbf{Relighting (Low)}} & \textbf{Chamfer L1} & \textbf{Time} \\ 
        
        & PSNR$\up$ & SSIM$\up$ & LPIPS$\down$ & PSNR$\up$ & SSIM$\up$ & LPIPS$\down$ & PSNR$\up$ & SSIM$\up$ & LPIPS$\down$ & Distance$\down$ & per step \\ \midrule
        SDF Conv. & 29.583 & 0.9361 & 0.0856 & 21.119 & 0.9205 & 0.0917 & 26.759 & 0.9140 & 0.0990 & 0.0080 & 10.85s \\
        SDF Reparam. & 33.141 & 0.9509 & 0.0586 & 26.092 & 0.9379 & 0.0616 & 29.039 & 0.9290 & 0.0744 & 0.0073 & 7.08s \\ 
        SDF Reparam. (hqq) & 29.621 & 0.9302 & 0.0943 & 22.262 & 0.9138 & 0.0996 & 30.385 & 0.9317 & 0.0723 & 0.0122 & 5.00s \\ \midrule
        Ours (trilinear) & \textbf{37.550} & \textbf{0.9812} & \textbf{0.0216} & 31.340 & 0.9722 & \textbf{0.0194} & 31.618 & 0.9574 & \textbf{0.0372} & 0.0052 & 2.23s \\
        Ours (tricubic) & 37.466 & 0.9800 & 0.0234 & \textbf{31.744} & \textbf{0.9729} & 0.0200 & \textbf{32.189} & \textbf{0.9582} & 0.0375 & \textbf{0.0047} & 4.03s \\
        \bottomrule
    \end{tabular}
    \caption{\textbf{Quantative evaluations}. We quantitatively measure the performance of different methods on synthetic Chair, Lego, Hotdog, Ficus, and Drum. For 2D evaluations, we test novel view rendering and relighting on a high-contrast and a low-contrast environment map. For 3D evaluations, we test the Chamfer L1 distance using random sample points on the ground truth mesh. We use the same optimization setup as in \autoref{fig:compare} and additionally run SDF Reparameterization using their released hqq setup. Since previous methods require tricubic interpolation of the SDF grid, we further test our method on both trilinear and tricubic interpolation for better reference.}
    \label{table:compare}
\end{table*}

\section{results}

We implemented our differentiable renderer using the \textit{Mitsuba3} \cite{Mitsuba3} Python package. Our implementation runs on CUDA and LLVM backends, and the results in this section were obtained using the CUDA backend on an NVIDIA RTX 3090.

\subsection{Validation}

\paragraph{Forward derivatives}

We validate our method by computing the forward derivative of the rendered image with respect to a single translation parameter. In Figure \ref{fig:derivative}, we place an object under a top area light and above a bottom plane and set the camera to look at the cast shadow of the object onto the bottom plane. In this way, all gradients come from the boundary integral. For reference, we use finite differences (FD) with a step size of $10^{-4}$. We also compare our results with SDF Convolution \cite{bangaru2020warpedsampling}, SDF reparameterization, and mesh projective sampling \cite{Zhang2023Projective}. Our method almost always results in the highest PSNR under different sizes of area lights. In Figure \ref{fig:mirror}, we place a metallic bunny in front of a mirror to produce inter-reflections. Again our differentiable renderer can accurately compute derivatives from higher-order light transport.


\paragraph{Inverse rendering}

We further test our differentiable direct integrator on end-to-end inverse rendering tasks. Note our focus is on demonstrating the effectiveness of our method on downstream applications, as this shows that the introduced bias is in practice small enough not to be a problem. We intentionally use a conservative setup and do not push for state-of-the-art performance.

In all our reconstructions, we use environment maps for lighting and perspective cameras distributed on the unit hemisphere. We restrict the SDF to be within the unit cube and use either diffuse or principled BSDF \cite{Burley2012PhysicallyBasedSA} as our material. We use voxel grids to represent the SDF and the albedo, roughness, and metallic parameters of the material. For optimization, we initialize with a sphere and run an Adam optimizer \cite{KingBa15} with a learning rate of $10^{-2}$. We adopt a coarse-to-fine optimization scheme that repeatedly upscales the grid resolution by $2$ after a fixed and sufficiently large number of iterations. Within each iteration, we compute the L1 loss between a batch of reference images and rendered images. To regularize the SDF to satisfy the eikonal constraint, we redistance the SDF after every iteration using the same fast sweeping method implementation of Vicini et. al. \cite{Vicini2022sdf}.

In \autoref{fig:optimize}, we demonstrate the optimization process on a variety of shapes. The high-genus Voronoi Bunny (24 views) showcases that we can handle complex geometry with frequent boundaries. Subsequently, we jointly optimize the geometry and the material of Lego, Chair, Hotdog, Ficus, and Drum (50 views) for more complete inverse rendering. We take the original Blender models of the NeRF Synthetic dataset \cite{mildenhall2020nerf} and adapt them to modified Mitsuba versions. Finally, to emphasize the physically-based nature of our differentiable renderer, we reconstruct the Logo (4 views) by only looking at its cast shadow onto the planes. We intentionally design the lighting condition so that the shadows largely overlap with each other, adding difficulty to the reconstruction.

\begin{figure*}[h!]
    \centering
    \includegraphics[width=\linewidth]{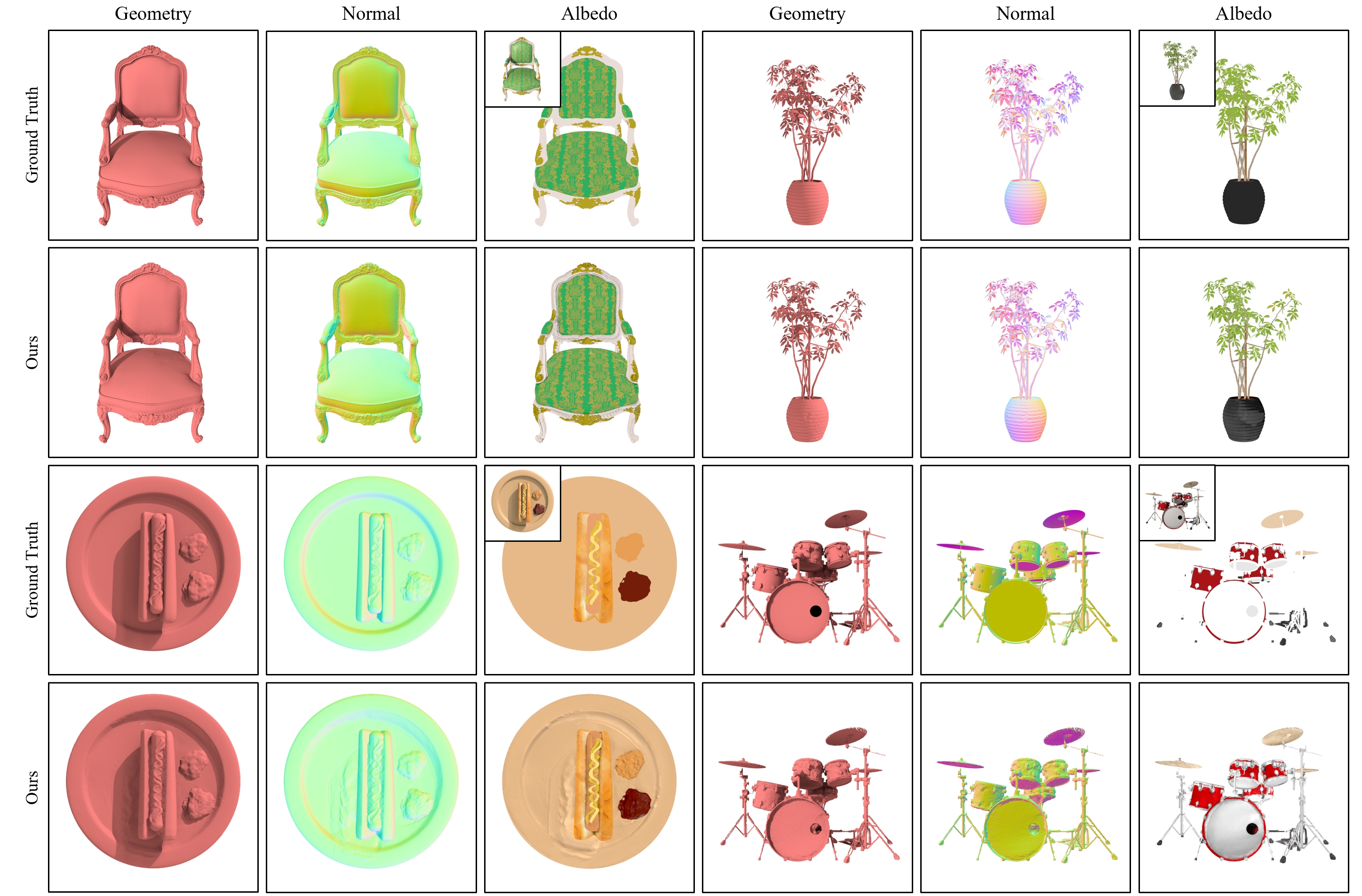}
    \includegraphics[width=\linewidth]{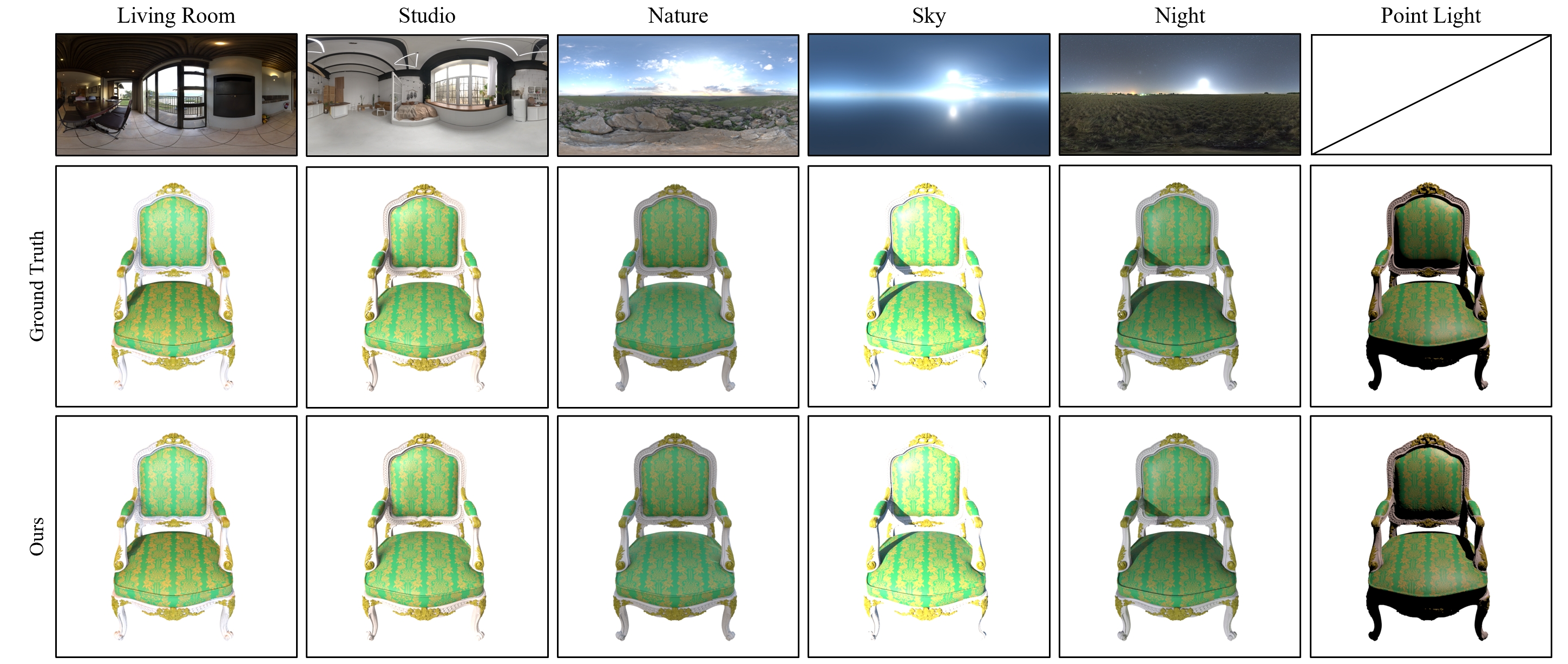}
    \caption{\textbf{Inverse rendering and relighting}. The physically-based nature of our differentiable renderer enables joint optimization of geometry/material and hence easy relighting. Here we visualize the geometry, normal, and albedo of our final inverse rendering and the relighting under various lighting conditions. Note how we can largely disentangle shading effects from the albedo (in comparison with the top left rendering).}
    \label{fig:inverse}
\end{figure*}

When it comes to inverse rendering, physically based differentiable renderers can disentangle the geometry and material. In \autoref{fig:inverse}, we visualize the geometry, normal, and albedo of our final inverse rendering results, as well as the relighting of the Chair under various lighting conditions. In general, we can correctly separate the geometry and material from the complex shading effects, such as the cast shadow on the Chair and the dense self-occlusions of the Ficus. We acknowledge that minor baking of the geometry into the albedo exists and the poor lighting conditions in shadowed area can lead to subpar results.


In \autoref{fig:compare}, we show a side-by-side comparison between SDF Convolution, SDF Reparameterization, and our method. In \autoref{table:compare}, we quantitatively measure the performance of different methods. For all tests on all shapes, we achieve results comparable or superior to the previous methods. We attribute this to effective sampling of the silhouette, which enables us to reconstruct fine structures like the stem of the Ficus and the belt on the Lego. This also helps reduce the gradient variance to achieve smoother surfaces on the Chair and Hotdog. Furthermore, gradient variance is important for improving robustness: extreme gradients can lead to too-large optimization steps and hence irremediable local blow-ups, as seen in the Chair and Lego reconstruction using SDF Convolution (red arrows).

\begin{figure}
    \centering
    \includegraphics[width=\linewidth]{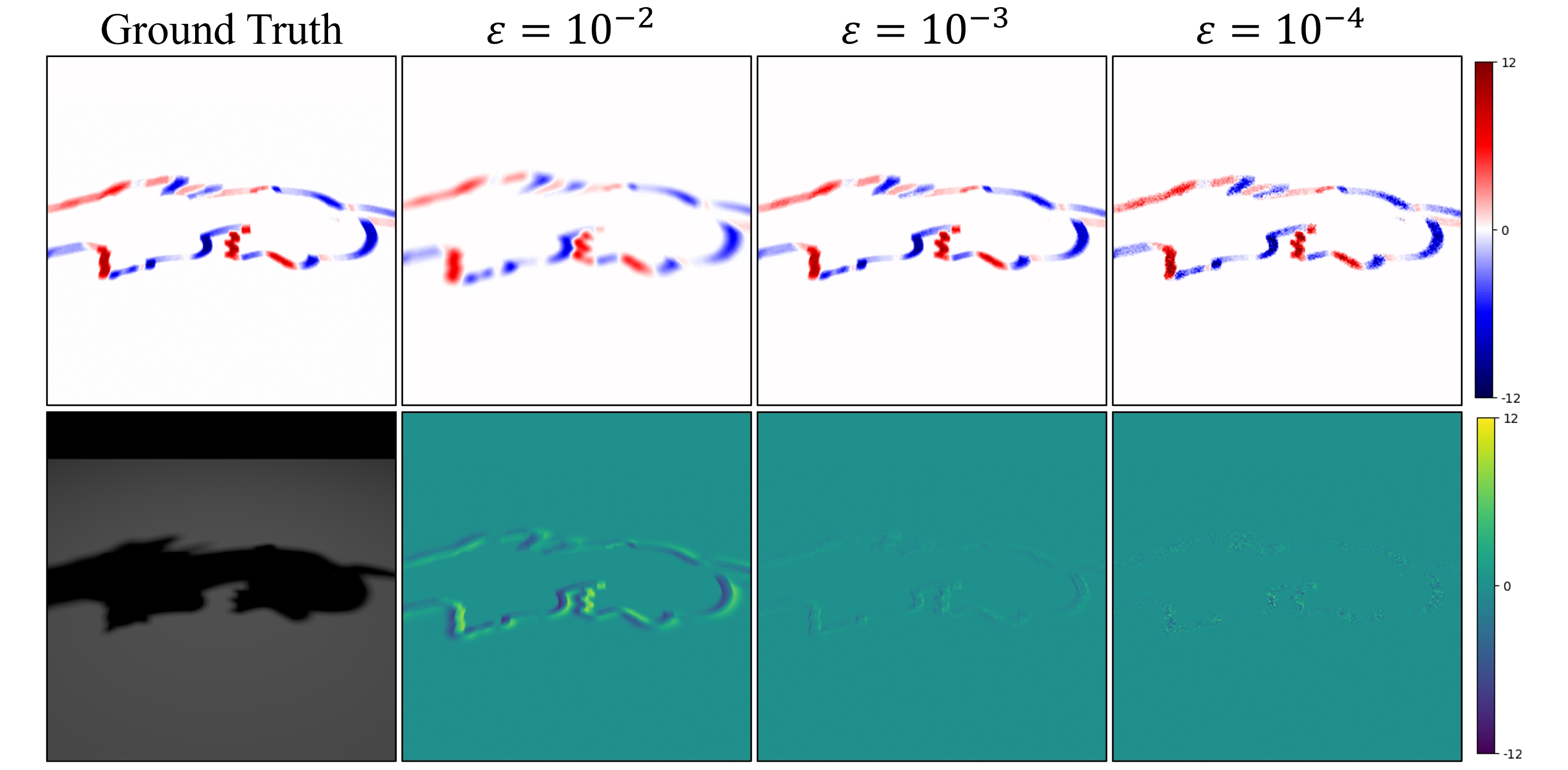}
    \caption{\textbf{Bias-variance tradeoff}. We show the effect of the SDF threshold on the gradients. This first row shows the reference and our forward derivatives, and the second row shows the rendered image and the difference between our forward derivatives and the reference. As $\epsilon$ gets lower, the gradients become noisier, but at the same time only Monte Carlo noise is observed in the difference. Here we compute our forward derivatives using $256$ spp.}
    \label{fig:bias-variance}
\end{figure}

\subsection{Sensitivity Analysis}

The SDF threshold $\epsilon$ is an important hyperparameter that controls how much we relax the visibility boundary. As $\epsilon$ increases, we smooth out the boundary more and the bias increases. However, higher $\epsilon$ enhances the probability of sampling relaxed silhouette points, so the gradients have less variance (Figure \ref{fig:bias-variance}) and it is not desirable to keep decreasing $\epsilon$. We refer to this phenomenon as the \textit{bias-variance tradeoff} of $\epsilon$. In all our reconstructions, where the target object is within a unit cube, we set $\epsilon = 10^{-4}$ to achieve the best of both worlds.

\begin{figure}
    \centering
    \includegraphics[width=\linewidth]{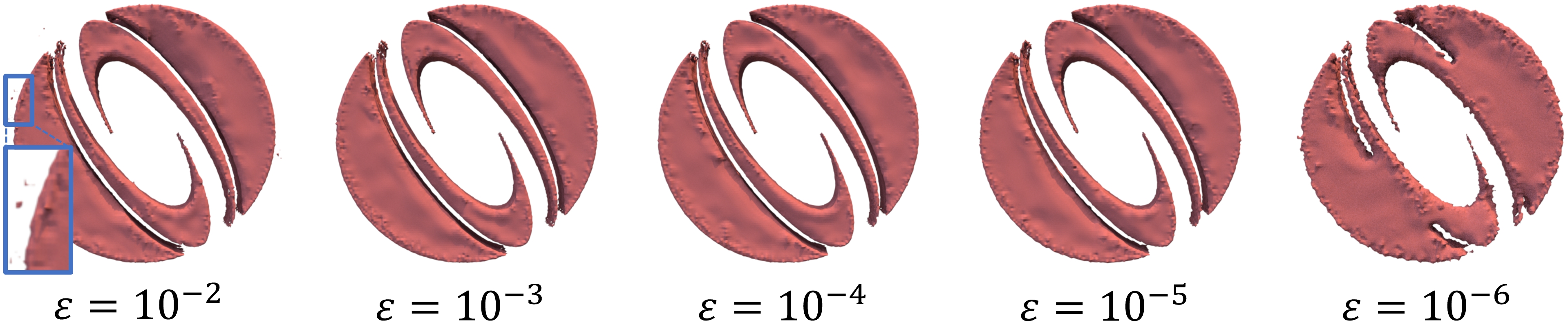}
    \caption{\textbf{Effect of $\epsilon$}. Our differentiable renderer works on a wide range of $\epsilon$. However, too large $\epsilon$ blurs the silhouette too much and causes floaters, while too little $\epsilon$ leads to insufficient sampling of the silhouette and causes slower convergence.}
    \label{fig:eps}
\end{figure}

\begin{figure*}
    \centering
    \includegraphics[width=\linewidth]{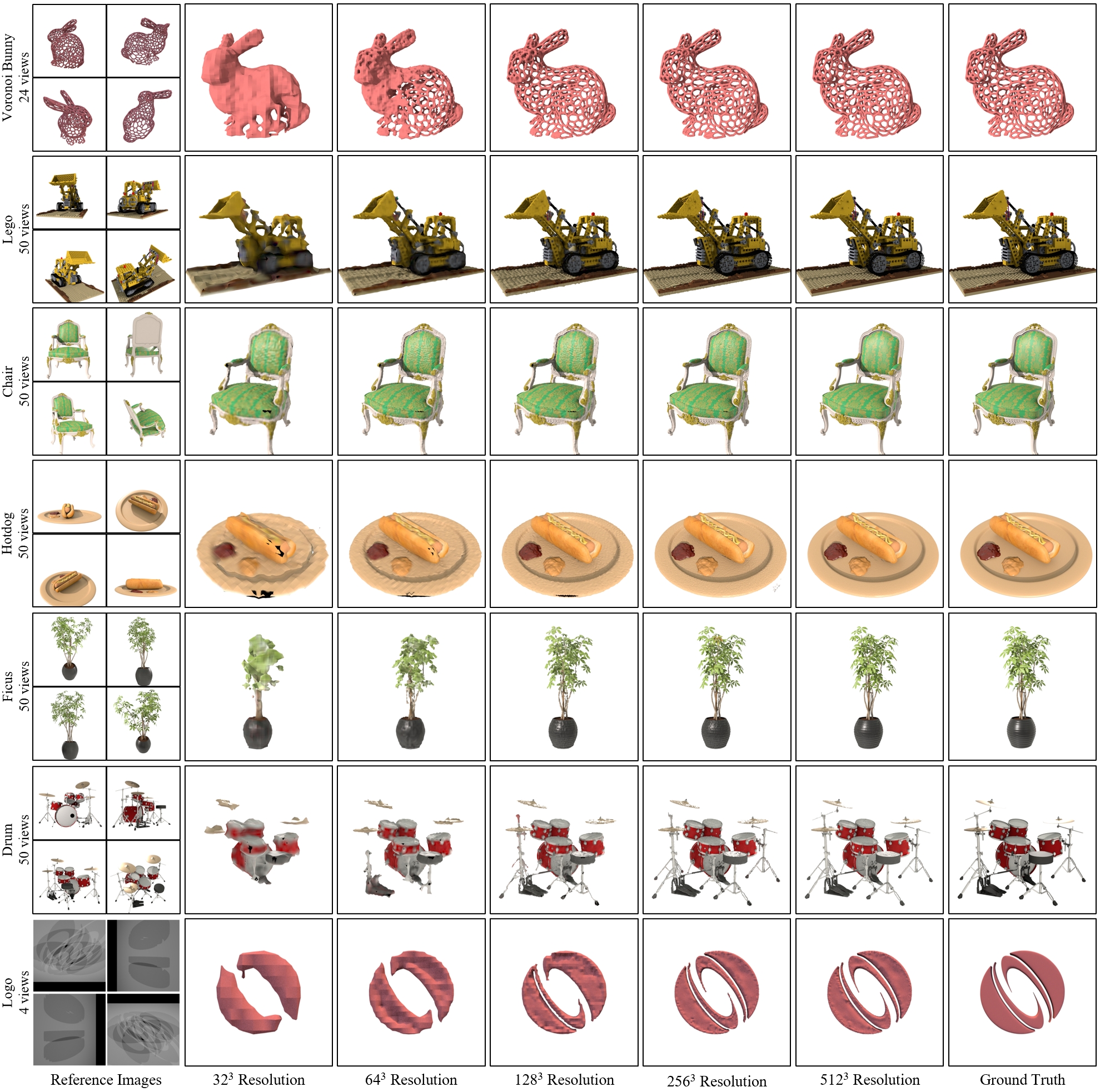}
    \caption{\textbf{Optimization Process}. We demonstrate the optimization process on a variety of shapes. For Voronoi Bunny, we only optimize the geometry. For Lego, we assume known diffuse material and jointly optimize the SDF and the albedo. For Chair, Hotdog, and Ficus, we optimize the SDF and a principled material (albedo and roughness). For Drum, we also add a metallic component. For Logo, we optimize the geometry by looking only at its cast shadow onto fixed walls. Voronoi Bunny from \cite{mehta2022level}. Lego $\copyright$Heinzelnisse. Drum $\copyright$bryanajones. Logo from \cite{Vicini2022sdf}.
    }
    \label{fig:optimize}
\end{figure*}

In addition to setting $\epsilon$ for optimizations inside the unit cube, we also want to know the feasible range of $\epsilon$. This is particularly important if we want our differentiable renderer to work on large scenes, where the distances of different objects to the camera can differ drastically. In \autoref{fig:eps}, we use the same shadow optimization setup of Logo as in \autoref{fig:optimize}, which helps rule out the influence of the interior integral. It turns out that our differentiable renderer works quite robustly with different scales of $\epsilon$. In all cases, our differentiable render succeeded in reconstructing the overall geometry. When $\epsilon = 10^{-2}$, too much blurring of the silhouette causes tiny floaters. When $\epsilon = 10^{-6}$, insufficient sampling of the silhouette causes the optimization not to fully converge. 


\subsection{Limitations}

\paragraph{Bias} The approximation of the silhouette through relaxed silhouette points is essentially biased. However, embracing nonzero bias, in turn, enables us to achieve architectural simplicity and efficient discontinuity sampling. We show through a series of experiments that the bias is well-controlled and does not interfere with downstream applications of our differentiable renderer.

\paragraph{Tuning SDF threshold} The optimal choice of the SDF threshold $\epsilon$ might depend on how far the objects are from the camera, which then might require tuning of $\epsilon$. For this reason, we show that our method is not very sensitive to $\epsilon$ and supports a wide range of feasible $\epsilon$. Users of our differentiable renderer should only need to tune the order of magnitude of $\epsilon$.

\paragraph{Solving for the relaxed silhouette point} Our relaxation scheme specifically asks for points within a certain distance of the surface. This means that we need to support the query of the distance of an arbitrary point to the surface, and hence we choose SDF as our surface representation. At the same time, this also means that our relaxed boundary integral can be extended to other representations, as long as there exists an efficient way to satisfy this query.

\section{Conclusion}

We present a novel differentiable rendering method of SDFs that is simple, robust, accurate, and efficient. Our relaxed boundary integral provides a new perspective to solving the long-standing silhouette sampling problem: through proper relaxation of the silhouette, we are able to use nearby Monte Carlo samples to directly approximate it. We test our differentiable renderer in downstream inverse rendering applications and achieve comparable or superior performance to previous methods.

For future work, one direction is to extend the relaxed boundary integral to support other shape representations. In addition, the constant $1/\epsilon$ factor in the relaxed boundary integral can be seen as a uniform weighting of all relaxed silhouette points. Investigating better weighting schemes to reduce bias would be another interesting open question.







\newpage
\bibliographystyle{ACM-Reference-Format}
\bibliography{reference}

\newpage


\end{document}